\pdfoutput=1
\documentclass[11pt]{article}
\usepackage[citecolor=blue]{hyperref}
\usepackage{amsmath,amsfonts,amssymb,amsthm}
\usepackage{mathrsfs}
\usepackage{verbatim}           
\usepackage{enumerate}
\usepackage[pdftex]{graphicx}
\usepackage[pdftex]{color}         
\usepackage{amscd}
\usepackage[footnotesize]{caption}

\numberwithin{equation}{section}

\title{Advancements in the ADAPT Photospheric Flux Transport Model}
\author{ Kyle S. Hickmann$^{\thanks{
             Applied Mathematics and Plasma Physics,
             Los Alamos National Laboratory, 
             hickmank@lanl.gov}}$, 
         Humberto C. Godinez$^{\thanks{
             Applied Mathematics and Plasma Physics,
             Los Alamos National Laboratory, 
             hgodinez@lanl.gov}}$,
         Carl J. Henney$^{\thanks{
             AFRL/Space Vehicles Directorate, 
             Kirtland AFB,
             cjhenney@gmail.com}}$,
         C. Nick Arge$^{\thanks{             
             AFRL/Space Vehicles Directorate, 
             Kirtland AFB,
             cnarge63@gmail.com}}$
}

\date{}
\begin{document}
\maketitle

\begin{abstract}
  Global maps of the solar photospheric magnetic flux are fundamental drivers
  for simulations of the corona and solar wind and therefore are important
  predictors of geoeffective events. However, observations of the solar
  photosphere are only made intermittently over approximately half of the
  solar surface. The Air Force Data Assimilative Photospheric Flux Transport
  (ADAPT) model uses localized ensemble Kalman filtering techniques to adjust
  a set of photospheric simulations to agree with the available
  observations. At the same time, this information is propagated to areas of
  the simulation that have not been observed. ADAPT implements a local
  ensemble transform Kalman filter (LETKF) to accomplish data assimilation,
  allowing the covariance structure of the flux transport model to influence
  assimilation of photosphere observations while eliminating spurious
  correlations between ensemble members arising from a limited ensemble
  size. We give a detailed account of the implementation of the LETKF into
  ADAPT. Advantages of the LETKF scheme over previously implemented
  assimilation methods are highlighted.
\end{abstract}


\medskip


\medskip


\section{Introduction}
The dynamic magnetic fields of the Sun, from coronal holes to strong
photospheric magnetic fields within active regions, play a central role in
driving Earth's space environment. The global solar photosphere represents the
boundary conditions for corona and solar wind models. Therefore, accurately
establishing the global solar photospheric magnetic field provides a better
understanding and specification of Earth's space weather environment.

Solar magnetic field measurements are recorded on a regular basis for half the
solar surface, \emph{e.g.} the National Solar Observatory (NSO) \emph{Global
  Oscillation Network Group} (GONG) \cite{HarveyEtAl1996} and the
\emph{Helioseismic and Magnetic Imager} (HMI): \cite{SchouEtAl2012} onboard
the \emph{Solar Dynamics Observatory} (SDO). Additional efforts are being made
to use helioseismic techniques to estimate active regions on the far side of
the Sun by analyzing acoustic waves propagating from the far side of the Sun
to the Earth side \cite{LindseyBraun1997}. The \emph{Solar Terrestrial
  Relations Observatory} (STEREO); \cite{SockerEtAl2000} mission from NASA
has provided $360^{\circ}$ observations from the solar corona, making it the
first ever data set that provides an almost complete nowcast (estimate of the
current state) of the solar corona. Additionally, methods of reconstructing
the Sun's coronal dynamics from observation have been advancing
\cite{butala2010dynamic}. Although many of these observational data sets
provide an accurate estimation of the current state of a portion of the solar
photosphere's magnetic flux, we lack the magnetograph observations needed to
generate an instantaneous map of the global photospheric magnetic flux
distribution.

Data assimilation methods are techniques that fuse information from
observational data into a physics-based model in order to align the model with
current physical conditions and improve forecasts
\cite{Daley1991,kalnay_atmospheric_2003}. The ensemble assimilation methods
discussed in this article are \emph{sequential}, as opposed to
\emph{variational}, and work by first randomly initializing several
realizations of a solar model; this is referred to as the
\emph{ensemble}. Next, the ensemble members are propagated forward in time
until a physical observation is made. Each ensemble member then contributes a
simulated observation and these observations are compared with the physical
observation. Lastly, the ensemble members are adjusted by weighting the
physical observation's noise and the ensemble of simulated observation's
variance. After the adjustment is performed, the \emph{analyzed} ensemble is
propagated forward in time again until the next physical observation is made
and the process is repeated. The large variety of data assimilation methods
specify the different forms for the adjustment that should be performed on the
ensemble.

Assimilation methods are starting to be implemented into solar modeling to
enable a better nowcast and forecast from a wide range of models
\cite{brun2007towards}. For example, \cite{butala2010dynamic} utilize the
ensemble Kalman filter along with a tomography method to reconstruct the
dynamic solar corona from real observational data and a random walk model of
coronal dynamics, \cite{kitiashvili2008application} also utilized the ensemble
Kalman filter to forecast the solar cycles and sunspot number by combining
real observations with a reduced $\alpha \Omega$-dynamo model, while
\cite{belanger2007predicting} used the four-dimensional variational method
(4D-Var) and a cellular-automaton-based avalanche model to predict simulated
solar flare data. A variational data assimilation technique was developed by
\cite{jouve2011assimilating} using an $\alpha \Omega$-dynamo model for the
Sun; their techniques were validated using synthetically generated
data. \cite{svedin2013data} applied a three-dimensional variational (3D-Var)
data assimilation methodology for a two-dimensional convection flow to
simulated observations. An ensemble Kalman filter method was implemented in a
Babcock--Leighton solar dynamo model, using synthetic data, by
\cite{dikpati2014}. Observations were combined with a physical model by
\cite{dikpati2006} who used and assimilation technique consisting of a simple
data-nudging method. The mentioned works illustrate just a few examples of a
wide range of problems from solar modeling that utilize data assimilation
methods. In this work, three different ensemble Kalman methods are compared as
data assimilation techniques for combining a physical model of the flux
transport dynamics of the solar photosphere and real vector spectral
magnetograph (VSM) observations. To the authors' knowledge, this represents
the first comparison of multiple sophisticated assimilation methods using a
physical solar photosphere model and real observations.

The Air Force Data Assimilative Photospheric Flux Transport (ADAPT) Model
\cite{arge2010air,arge2011improving,henney2012forecasting,arge2013modeling}
incorporates various data assimilation techniques, using a module developed at
Los Alamos National Laboratory (LANL) \cite{arge2010air}, with a photospheric
magnetic flux transport model. The ADAPT model is updated with observations by
using an ensemble of simulations from a flux transport model based on the
method used in the Worden--Harvey (WH) model \cite{worden2000evolving} to
represent the distribution of possible solar photospheric states under the
influence of differential rotation, meridional flow, supergranular diffusion,
and random emergence of weak background flux. The ensemble information is then
used in LANL's ensemble Kalman filter (ENKF) data assimilation method to
adjust the ADAPT model using observational data. The adjustment is made by
comparing the covariance structure of the model, as computed from the ensemble
samples, with an observation and its noise. In areas where the ensemble shows
a diffuse distribution, representing lack of determination in the model,
observations have a greater impact on the ensemble. Several types of Kalman
filtering techniques are now available for use with the ADAPT model. To date
all data assimilation methods implemented with ADAPT are \emph{sequential} as
opposed to \emph{variational} methods. Specifically the most recent LANL
module's data assimilation method in ADAPT implements a localized version of
ensemble Kalman filtering, which assimilates each observation within a local
region of the model spatial field. This localization allows approximation of
the ADAPT spatial covariance structure to effect the assimilation of new
observations more effectively. At the same time the localization in the data
assimilation eliminates erroneous long distance correlations occurring from
the small sample sizes used to estimate the ADAPT model's covariance
structure. It is worth noting that all ADAPT maps and results to date have
utilized an ensemble least squares (ENLS) data assimilation method
\cite{arge2011improving,henney2012forecasting}, whereas the ENKF data
assimilation methods developed at Los Alamos have only been used so far in a
testing and development capacity.

The article is outlined as follows: In Section \ref{methods} we give a brief
description of the flux transport method within the ADAPT model and detail the
implementation of the data assimilation scheme. This is followed by a
comparison of different data assimilation methods available in ADAPT in
Section \ref{DAcomparison}. The article concludes, in Section
\ref{discussion}, with a discussion of future directions for the ADAPT model.

\section{Methods}\label{methods}

\subsection{Flux Transport within ADAPT}

The photospheric flux transport within the ADAPT model is based on the
Worden--Harvey (WH) model \cite{worden2000evolving}, which includes the effects
of differential rotation, meridional flow, supergranulation, and random
background flux. New flux emergence is added to the ADAPT simulation through
data assimilation of observations. Mechanisms for the creation and destruction
of large scale active regions remains an important research topic
\cite{brun2007global,archontis2008magnetic,zhang2010statistical} but a
definitive model has yet to emerge.

The rate of rotation of the photospheric flux is dependent on latitude. The
sidereal rotation rate, at latitude $\theta$, is given by
\begin{equation}\label{differential_rotation_profile}
  \omega(\theta) = A + B \sin^2(\theta) + C \sin^4(\theta).
\end{equation}
The original WH model used parameter values from Snodgrass
\cite{snodgrass1983magnetic}, i.e. $A = 2.902 \, \mu \textrm{rad}/\mathrm{s}$, $B = -0.464 \, \mu \textrm{rad}/\mathrm{s}$, $C = -0.328 \, \mu
\textrm{rad}/\mathrm{s}$. Currently the ADAPT model uses parameter values
from \cite{komm1993rotation}, $A = 2.913 \, \mu \textrm{rad}/ \mathrm{s}$, $B
= -0.405 \, \mu \textrm{rad}/ \mathrm{s}$, $C = -0.422 \, \mu \textrm{rad}
/ \mathrm{s}$.

ADAPT also includes a (mostly poleward) meridional flow that is more difficult
to estimate from observations than is the nearly steady surface differential
rotation. Difficulty in meridional flow estimation is due in large part to the
flow rate being slow enough that it is difficult to distinguish from other
transport processes such as supergranular flows. Moreover, there is evidence
of significant meridional flow rate variation over the course of the solar
cycle \cite{hathaway2010variations}. The ADAPT model uses the profile, based
on \cite{wang1994rotation}, implemented in the original WH model given by
\begin{equation}\label{meridional_flow_profile}
M(\theta) = (8 \, m / \mathrm{s}) |\sin (\theta)|^{0.3} |\cos (\theta)|^{0.1}.
\end{equation}
This profile does not match observations, especially at lower latitudes,
however, with regular observational data assimilated into ADAPT only the high
latitudes ($|\theta| > 65^{\circ}$) are affected by Equation
(\ref{meridional_flow_profile}).

The diffusion of flux is approximated with supergranular flows. ADAPT uses a
stochastic diffusion method developed by
\cite{mosher1977magnetic} and \cite{simon1995kinematic} to simulate the diffusion and
transport of supergranules. The stochastic diffusion used by ADAPT is based on
the description in \cite{worden2000evolving}. This representation of
supergranular diffusion overestimates dissipation
\cite{worden2000evolving,mackay2012sun}, so the supergranular diffusion
process is shut off in ADAPT for field strengths greater than $15$ to $50$ G
depending on the data source and the rate of assimilation.

As noted in the original WH model, \cite{schrijver2001formation} point out
that the photospheric magnetic flux would disappear due to random
cancellations in two to three days if the total magnetic flux were not renewed
regularly. For this reason the ADAPT framework includes daily random
background flux emergence. This is accomplished by taking Gaussian distributed
random flux at each pixel with mean zero and absolute mean value of $2.1$ G in
each day of the simulation. This value of random flux emergence maintains a
constant level of total flux in the synoptic map.

\subsection{LETKF Data Assimilation}

We have implemented a local ensemble Kalman filter (LETKF) for testing and
development within ADAPT. Currently all ADAPT maps generated to date for solar
prediction research purposes have utilized the ensemble least squares (ENLS)
data assimilation method \cite{arge2011improving}. Our treatment of the LETKF
follows the work of Hunt \cite{hunt2007efficient}. The key difference between
the standard ensemble Kalman filter and the local Kalman filter is in the
treatment of the ensemble covariance. In all ensemble Kalman methods, the
model covariance is approximated from the individual ensemble members. Due to
the small ensemble sizes often allowed computationally, this covariance
approximation is very error prone and non-physical long distance correlations
can arise due to sampling error. The local Kalman filter alleviates this
problem by only calculating covariances in a region spatially localized around
the pixel being updated, therefore eliminating the influence of long distance
spurious correlations. We now describe the steps in the LETKF and the
localization particular to ADAPT.

\subsubsection{Transform} 

In the LETKF we denote each realization of the forecast synoptic map from the
ADAPT model as a column vector $x_{\mathrm{f}}$. Here the pixel values of the
synoptic map are taken as entries of the column vector. We denote the ensemble
mean of these column vectors by $\bar{x}_{\mathrm{f}}$. The forecast ensemble
matrix $X_{\mathrm{f}}$ is then formed with columns consisting of the
discrepancy vector between an ADAPT realization and the ensemble mean
$x_{\mathrm{f}} - \bar{x}_{\mathrm{f}}$. For each pixel in the rows of
$X_{\mathrm{f}}$, we define a set of \emph{local} observations, described
below. The observation operator is applied to each of the ensemble members to
form an ensemble of observations. This ensemble of observations has members,
denoted $y_{\mathrm{f}}$, with mean $\bar{y}_{\mathrm{f}}$. The ensemble
observation matrix $Y_{\mathrm{f}}$ is then formed with columns consisting of
the discrepancy between an ensemble observation and the ensemble observation
mean $y_{\mathrm{f}} - \bar{y}_{\mathrm{f}}$.

The transformation in the local ensemble transform Kalman filter is to view
the realizations and local observations of the ADAPT model as Gaussian random
variables \cite{hunt2007efficient}. For an ensemble of size $k$ if $\omega
\sim \mathcal{N}(\mathbf{0}, (k-1)^{-1}\mathcal{I})$ then
$\bar{x}_{\mathrm{f}} + X_{\mathrm{f}} \omega \sim
\mathcal{N}(\bar{x}_{\mathrm{f}}, (k-1)^{-1}X^T_{\mathrm{f}} X_{\mathrm{f}})$
similarly $\bar{y}_{\mathrm{f}} + Y_{\mathrm{f}} \omega \sim
\mathcal{N}(\bar{y}_{\mathrm{f}}, (k-1)^{-1}Y^T_{\mathrm{f}}
Y_{\mathrm{f}})$. These Gaussian random variables preserve the mean and
covariance structure of the original ensemble and ensemble observations as
sampled from the ADAPT model. Now the LETKF performs data assimilation in
$\omega$-space using $Y_{\mathrm{f}}$ as the observation operator. After
assimilation has been performed, it is easy to transform back to the ADAPT
ensemble space through multiplication by $X_{\mathrm{f}}$.

\subsubsection{Inflation} 

The photospheric observations often fall far enough away from the entire ADAPT
ensemble that, even with localization of the ETKF data assimilation scheme,
observations are discarded by ADAPT and simulations diverge from observations
\cite{wang2003comparison}. Moreover, as the ADAPT ensemble of simulations
diverge, the ensemble variance will become smaller as observations are
continually discarded and the individual ensemble members are brought more in
agreement with each other. This latter phenomenon is known as \emph{ensemble
  collapse}. To remedy these problems, we implement inflation of the
ensemble. This artificially spreads out the ADAPT ensemble to envelope a wider
range of possible photospheric maps thereby increasing the likelihood of a
portion of the ensemble members being near the observations. We adjust each
forecast ensemble member and observation ensemble member using the inflation
factor $\rho$ and the transformation
\begin{align}\label{ensemble_inflation}
\tilde{x}_{\mathrm{f}} &= \bar{x}_{\mathrm{f}} + \rho (x_{\mathrm{f}} - \bar{x}_{\mathrm{f}}) \\
\tilde{y}_{\mathrm{f}} &= \bar{y}_{\mathrm{f}} + \rho (y_{\mathrm{f}} - \bar{y}_{\mathrm{f}})
\end{align}
for $\rho > 0$ \cite{kalnay_atmospheric_2003,evensen_data_2009}. The data
assimilation is then performed using this adjusted set of the ADAPT
realizations which has the effect of giving more weight to the
observations. The larger the inflation factor $\rho$ is chosen, the more
observations are favored over model forecasts
\cite{kalnay_atmospheric_2003,evensen_data_2009}. A careful choice of $\rho$
is important to balance the weight given to the observations and ADAPT
ensemble. Currently the choice of inflation factor is performed by trial and
error on historical observations, however automatic methods exist to choose
$\rho$ and these will be implemented into ADAPT in the future. In instances
when the inflation is large along with observational noise, localized ensemble
divergence can still develop (see Figure \ref{fig:stddev_image_comparison}).

\subsubsection{Analysis Ensemble} 

The actual photosphere observations being assimilated are denoted by
$y_{\textrm{obs}}$ and the observational error or noise will be assumed to
have covariance matrix $R$. In $\omega$-space the analysis mean and covariance
is then given by the usual Kalman update equations
\cite{kalnay_atmospheric_2003,evensen_data_2009}
\begin{align}\label{w_KF_analysis}
\bar{\omega}_a &= \tilde{P}_a Y^T_{\mathrm{f}} R^{-1} (y_{\textrm{obs}} - \bar{y}_{\mathrm{f}}) \\
\tilde{P}_a &= [(k - 1) \mathcal{I} + Y^T_{\mathrm{f}} R^{-1} Y_{\mathrm{f}}]^{-1},
\end{align}
where $k$ is the number of members within the ADAPT ensemble and $R$ is the
observational error covariance matrix described below.

The mean $\bar{\omega}_a$ and covariance $\tilde{P}_a$ are transformed to the
ADAPT ensemble space using $X_{\mathrm{f}}$ as an operator. Thus, the analysis
mean $\bar{x}_a$ and covariance $P_a$ in ensemble space are
\cite{hunt2007efficient}
\begin{align}\label{X_KF_analysis1}
\bar{x}_a &= \bar{x}_{\mathrm{f}} + X_{\mathrm{f}} \bar{\omega}_a \\ \label{X_KF_analysis2}
P_a &= X_{\mathrm{f}} \tilde{P}_a X^T_{\mathrm{f}}.
\end{align}

Equations (\ref{X_KF_analysis1}) and (\ref{X_KF_analysis2}) determine the mean
and covariance of ADAPT's analyzed ensemble.  However, one then must specify
the updated analysis ensemble members, denoted $x_a$. The updated analysis
ensemble members $x_a$ must have mean and sample covariance that satisfy
Equations (\ref{X_KF_analysis1}) and (\ref{X_KF_analysis2}). ADAPT uses the
square root filter method to ensure the ensemble members are updated in such a
way that the analysis ensemble has mean $\bar{x}_a$ and covariance
$\tilde{P}_a$
\cite{bishop2001adaptive,kalnay_atmospheric_2003,evensen_data_2009}. Namely,
we set the analysis ensemble in $\omega$-space to be
\begin{equation}\label{w_analysis_ensemble}
\Omega_a = [(k - 1) \tilde{P}_a]^{\frac{1}{2}}
\end{equation}
so the analysis ensemble matrix in ADAPT becomes $X_a = X_{\mathrm{f}}
\Omega_a$. The individual analysis ensemble members $x^{(i)}_a$ are then
formed using each column of $\Omega_a$,
\begin{equation}\label{x_analysis_ensemble}
x^{(i)}_a = \bar{x}_a + X_{\mathrm{f}} \Omega^{(i)}_a, i = 1, 2, \dots, k.
\end{equation}
The square root in Equation (\ref{w_analysis_ensemble}) is the symmetric
square root obtained through the singular value decomposition of $(k-1)
\tilde{P}_a$, as opposed to the matrix square root obtained through the
Cholesky decomposition. Using the symmetric square root is necessary to
preserve continuity during localization
\cite{bishop2001adaptive,wang2003comparison,hunt2007efficient}.

\subsubsection{Photosphere Localization} 

In the above analysis scheme, it is possible to perform the data assimilation
one pixel at a time by taking the forecast ensemble matrix $X_{\mathrm{f}}$ to be a row
vector of the ensemble discrepancies at a single pixel. One can then iterate
over all the pixels in the ADAPT forecast to generate an analysis ensemble.

For the assimilation of an individual pixel, one can either use all of the
observed pixels on the Earth side to form the observation ensemble discrepancy
matrix, $Y_{\mathrm{f}}$, or only the observed pixels that are believed to be
highly spatially correlated due to properties of the flux transport model or
spatial coordinate system
\cite{houtekamer2001sequential,hamill2001distance,whitaker2002ensemble,ott2002exploiting,ott2004local}. Including
only the observed pixels highly correlated with the pixel value being analyzed
reduces the effect of spurious correlations that arise in the observation
ensemble due to the small ensemble sizes
\cite{houtekamer2001sequential,hamill2001distance,whitaker2002ensemble,ott2002exploiting,ott2004local}.

In ADAPT, the longitudinal coordinate system of the solar photosphere causes
centers of pixels near the Equator to be much farther apart than centers of
pixels near the Poles. The spatial distortion caused by the longitudinal
coordinate gives a natural local region centered on each pixel that is highly
correlated with that pixel's current value. The selected localization region
has an ellipsoidal shape with axes aligned with solar longitude and
latitude. Pixel centers are much closer together at the Poles than close to
the Equator, and therefore we hypothesize that the correlation between pixel
values decreases more slowly as a function of longitudinal distance near the
Poles. The dependence of longitudinal correlation on latitude motivates us to
define our local ellipse to have a constant radius in the latitudinal
direction and a longitudinal radius that increases away from the Equator.

To describe the local observation region, let the $(i,j)$ synoptic pixel value
of an ensemble member be denoted by $x^{ij}_{\mathrm{f}}$. The forecast
ensemble matrix for this pixel is the row vector $X^{ij}_{\mathrm{f}}$ made up
of the different ensemble members $(i,j)$ pixel values minus their
average. The $(i,j)$ pixel has a corresponding solar latitude and longitude
$(\theta_i, \phi_j) \in \left[ -\frac{\pi}{2}, \frac{\pi}{2} \right] \times
[0, 2\pi)$. For each synoptic pixel value we define a local region of
observation $\mathcal{O}_{ij}$ based on the location $(\theta_i,
\phi_j)$. During the analysis computation any pixel location falling inside
$\mathcal{O}_{ij}$ contributes to the columns of the local observation
ensemble matrix $Y^{\textrm{loc}}_{\mathrm{f}}$. Any observations with
locations in $\mathcal{O}_{ij}$ make up the local observation vector
$y^{\textrm{loc}}_{\textrm{obs}}$ for the $(i,j)$ pixel. Now Equations
(\ref{w_KF_analysis})\,--\,(\ref{w_analysis_ensemble}) can be used with the
localized ensemble and observations to compute the analysis ensemble for the
$(i,j)$ pixel value.

ADAPT's LETKF data assimilation module sets the local observation region
$\mathcal{O}_{ij}$ to an ellipse with its major and minor axes aligned with
latitude and longitude. The ellipse's latitudinal radius is fixed at
$r_{\theta} = \frac{\pi}{60} = 3^{\circ}$ and the longitudinal radius
$r_{\phi}(\theta)$ is dependent on the latitude from the solar Equator. Since
the longitudinal coordinate causes correlations over shorter longitudinal
distances near the Equator, $r_{\phi}(\theta)$ is set to reach its maximum at
$\theta = 85^{\circ}$ and increase linearly as the latitude approaches the
Poles. We set
\begin{equation}\label{loc_long_radius}
r_{\phi}(\theta) = 3^{\circ} + 12^{\circ} \frac{|\theta^{\circ}|}{85^{\circ}} = \frac{\pi}{60} - \frac{12}{85} | \theta |
\end{equation}
and the local ellipse becomes
\begin{equation}\label{local_ellipse}
\mathcal{O}_{ij} = \left\{ (\theta, \phi) : \frac{(\theta - \theta_i)^2}{r^2_{\theta}} + \frac{(\phi-\phi_j)^2}{r^2_{\phi}(\theta_i)} < 1 \right\}.
\end{equation}

\subsubsection{Observation Covariance} 

The observational error covariance matrix $R$ is specified through
photospheric observations. Only the observation standard deviation is
specified at each pixel, so we assume that $R$ is diagonal and the
observational noise is not spatially correlated. The observational inputs
utilized by ADAPT are from line-of-sight magnetogram data from the Kitt Peak
Vacuum Telescope Vector Spectromagnetograph (VSM)
\cite{henney2007solis}. ADAPT generates a new map each time an observed
magnetogram is available. Typically, the VSM full-disk magnetograms are
available at a cadence of approximately one per day. Magnetograph data from
additional instruments can be used with ADAPT, with the caveat that the
inferred photospheric field strengths between instruments can vary by as much
as a factor of two \cite{riley2007}. The estimated observational error is
3\,\% with a sharp increase towards the limb to give more weight to the model
values.

\section{Data Assimilation Comparison}\label{DAcomparison}

The main difference between the LETKF data assimilation implementation and the
older \cite{arge2010air,arge2011improving,arge2013modeling} ENLS data
assimilation schemes used is how much the ADAPT ensemble is adjusted to agree
with the observations. With the ENLS data assimilation approach, the spatial
correlation structure of the ADAPT ensemble arising from the flux transport
model is not taken into account. This causes observations to be trusted far
more than the ADAPT model forecast, and therefore the ADAPT forecast is nearly
discarded during the ENLS assimilation. In sections of the observation region
near the central meridian, where observational noise is low, this can be
acceptable. However, near the limbs of the observation region, noise is
considerable and discarding the model forecast is not desirable.

Some of the spatial covariance structure in ADAPT's flux transport model is
included when using the non-localized Kalman filtering module. However, we
will show that a pure implementation of the ensemble transform Kalman filter
(ETKF) has many drawbacks due to spurious correlations introduced through
small ensemble sample size. Localization of the ETKF alleviates these spurious
correlations and provides a useful compromise between the ENLS and ETKF
methods. We will show how ADAPT with the standard ensemble transform Kalman
filter restricts the variance away from observations too much, severely
reducing the variance in the ADAPT ensemble. This causes ensemble collapse
that effectively eliminates the assimilation of observations.

\begin{figure}[!h]
  \centerline{\includegraphics[scale=0.36]{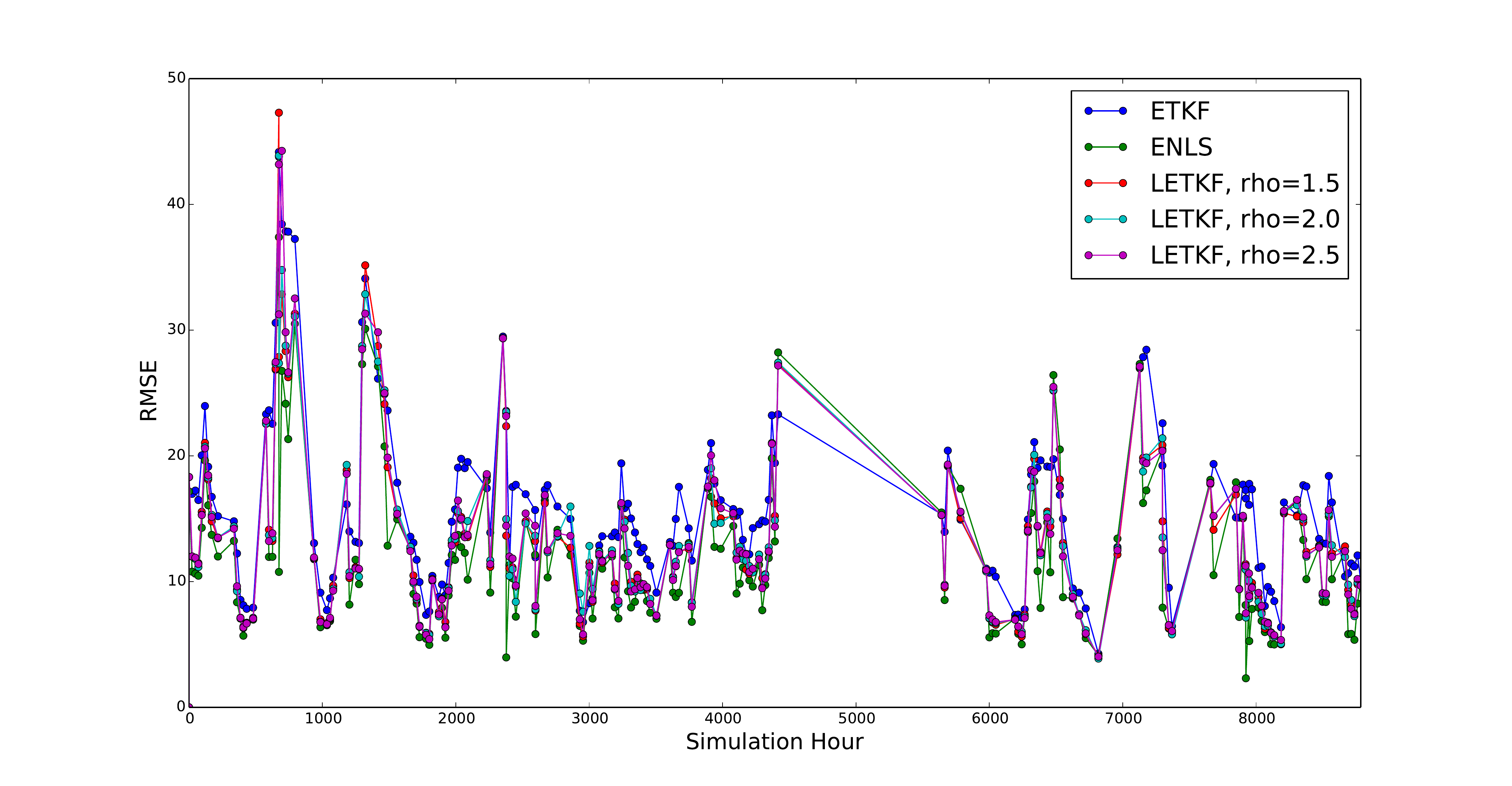}}
  \caption{Forecast RMSE over observation region: Here we show the time series
    of the average root mean square error (RMSE) between the forecast mean and
    the observation using each of the data assimilation methods available in
    ADAPT. The time period for this time series is between 26 September 2003
    and 26 September 2004. The average is taken over each pixel in the
    observation region. This figure shows that, in terms of RMSE, none of the
    data assimilation methods outperforms the others in the RMSE metric
    consistently over the period studied. However, RMSE is not the only method
    of distinguishing the usefulness of the data assimilation methods.}
  \label{fig:forecast_RMSE_comparison}
\end{figure}

To evaluate performance of multiple data assimilation methods researchers
often use a root mean square error (RMSE) approach. The RMSE is calculated by
taking the squared difference of the mean ensemble value before assimilation
and the current observation at each pixel. These squared differences are then
averaged over the observation region and the square root of the result is the
RMSE. A comparison of the RMSE time series for different ADAPT data
assimilation schemes is shown in Figure
\ref{fig:forecast_RMSE_comparison}. One can see that one method does not
outperform the others, in terms of RMSE, $100\%$ of the time. However, RMSE
does not account for how much one data assimilation method preserves the
physical model after adjustment. We argue below that this is the main
advantage of using the LETKF for ADAPT's data assimilation.

\begin{figure}[!h]
  \centerline{\includegraphics[scale=0.38]{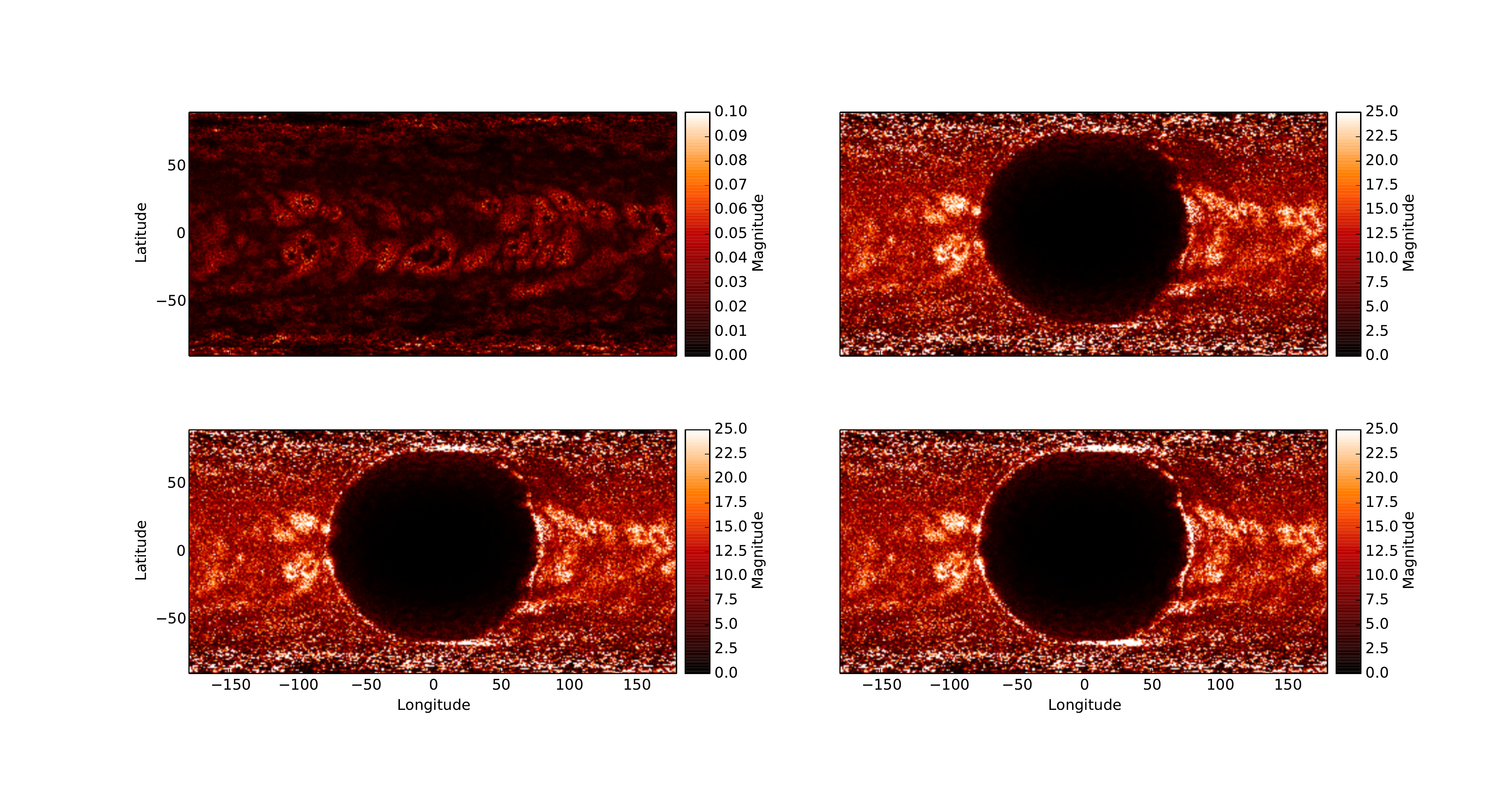}}
  \caption{ADAPT Standard Deviation Frames: In each of the above frames the
    pixel-by-pixel standard deviation of the analysis ensemble for different
    ensemble Kalman filtering methods is shown; ETKF (top left), LETKF with
    $\rho = 1.5$ (top right), LETKF with $\rho = 2.0$ (bottom left), LETKF
    with $\rho = 2.5$ (bottom right). All frames represent the same time in an
    ADAPT assimilation run using the same SOLIS-VSM observations. Namely, the
    data being assimilated were observed from SOLIS-VSM on 26 September 2003
    at 16:51. We can see the drastic reduction in pixel-wise ensemble variance
    under the ETKF scheme. Also, an artifact of the ensemble inflation under
    the LETKF scheme is noticeable. Near the polar regions, observation noise
    (variance) is high. This causes the restriction of the pixel-wise ensemble
    variance under LETKF to be slight near the polar regions. Thus, the
    standard deviation blows up for $\rho = 2.0$ and $\rho = 2.5$ near these
    regions. This shows up as a white streak near the polar portion of the
    observation region in the bottom frames.}
  \label{fig:stddev_image_comparison}
\end{figure}

\subsection{ETKF vs. LETKF} 

In situations, such as in solar photosphere models, where the dimension of the
simulation state space is high, small ensemble size will give rise to spurious
correlations
\cite{houtekamer2001sequential,hamill2001distance,whitaker2002ensemble,ott2002exploiting,ott2004local}. In
the case of the solar photosphere these occur over long distances and thus
severely restrict the analysis ensemble's pixel-wise standard deviation both
near and far away from observations (see Figure
\ref{fig:stddev_image_comparison}). On the other hand, the local ensemble
transform Kalman filter (LETKF) only compares each pixel's observation with a
model ensemble of pixels nearby, as described above. This eliminates the
propagation of strong correlations over long distances due to the small
ensemble size. We can see in Figure \ref{fig:stddev_image_comparison} that the
pixel-wise standard deviation is only severely restricted in the interior of
the observation region.

\begin{figure}[!h]
  \centerline{\includegraphics[scale=0.45]{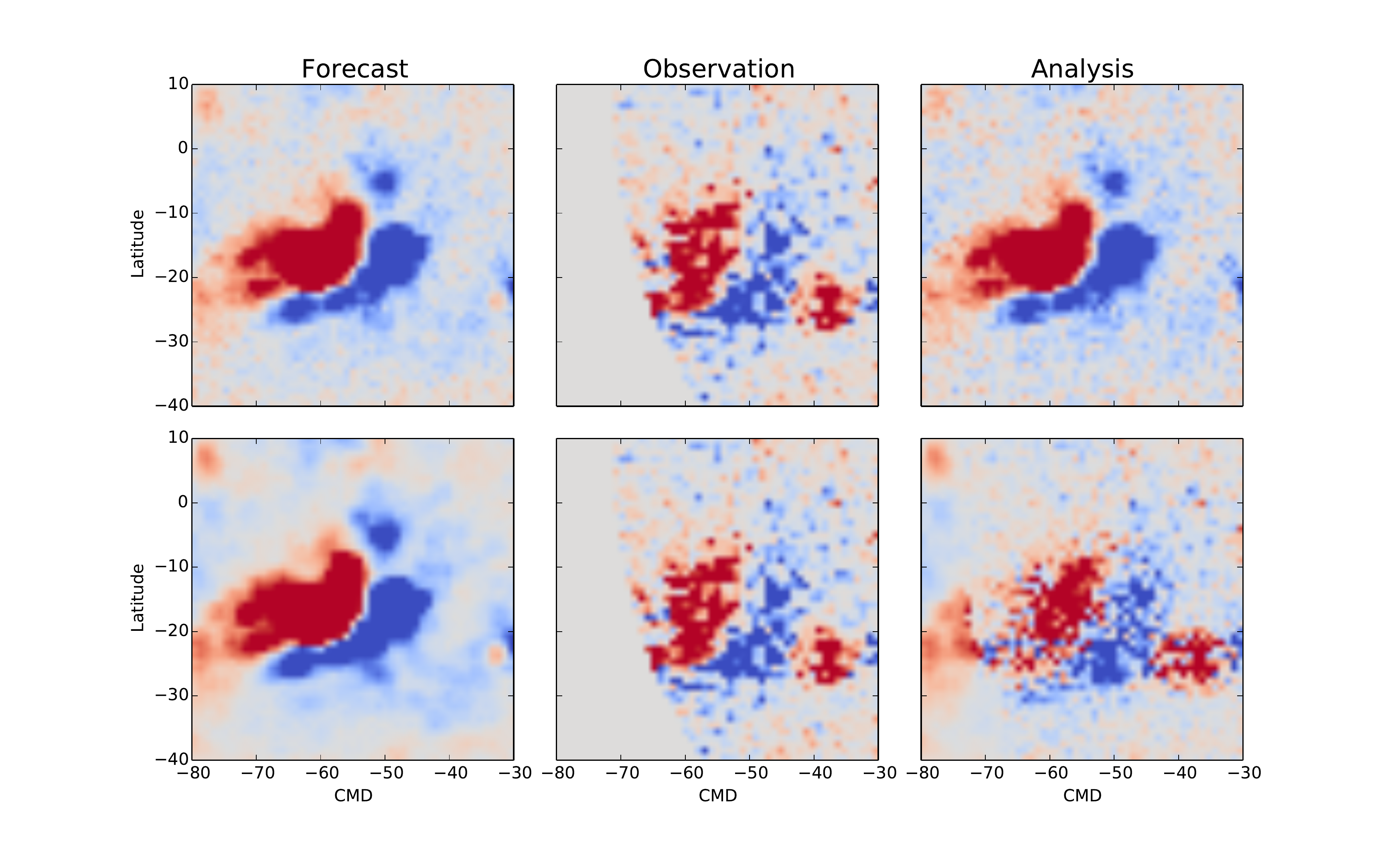}}
  \caption{Comparison of ETKF and LETKF assimilation effects: The top row
    represents one step of the ETKF assimilation. The Bottom row is one step
    of the LETKF assimilation. The first column is the mean forecast, second
    column is the observation, and the last column is the analyzed ADAPT
    ensemble mean. We can see that the ETKF essentially ignores the
    observation while the LETKF blends the observation and forecast
    ensemble. The data being assimilated were observed from SOLIS-VSM on 26
    September 2003 at 16:51.}
  \label{fig:etkfVSletkf_activeregion}
\end{figure}

The main effect of the variance reduction, in terms of accuracy of the data
assimilation, is how much the observations are taken into account when
adjusting the photosphere ensemble. The contrast is highlighted by observing
one assimilation step for a large active region using the two methods, as seen
in Figure \ref{fig:etkfVSletkf_activeregion}. The mean shape of the active
region is almost unaffected by the observations for the ETKF but is noticeably
influenced by observations when the LETKF is utilized.

\subsection{ENLS vs. LETKF} 

The ensemble least squares (ENLS) data assimilation that ADAPT has used in the
past \cite{arge2011improving} suffers from the opposite problems to those
that hinder the ETKF. Briefly, the ENLS method \cite{bouttier2002data}
updates each ensemble member pixel-by-pixel using the formula
\begin{equation} 
x_a = x_{\mathrm{f}} + \frac{\sigma_{\mathrm{f}}^2}{\sigma_{\mathrm{f}}^2 + \sigma_{\textrm{obs}}^2} (y_{\textrm{obs}} - x_{\mathrm{f}}),
\end{equation}
where $x_{\mathrm{f}}$ is a pixel value of the ensemble member being updated,
$y_{\textrm{obs}}$ is the observed value for that pixel, $\sigma_{\mathrm{f}}$
is the ensemble standard deviation for the pixel, $\sigma_{\textrm{obs}}$ is
the observation noise standard deviation, and $x_a$ is the analyzed ensemble
pixel value. In the ENLS scheme a pixel in the ensemble is only updated if
there exists a direct observation of its value. With the ENLS, observations
are assimilated into the ADAPT ensemble pixel by pixel without taking sampled
spatial correlations into account. Only pixel-wise standard deviations are
considered, resulting in local distortion of coherent structures, such as
large active regions, in the photospheric magnetic flux present in the
ensemble. This is due to noise in magnetic flux observations that is not
spatially correlated and therefore reduces spatial correlations in the
observation.

\begin{figure}[!h]
  \centerline{\includegraphics[scale=0.47]{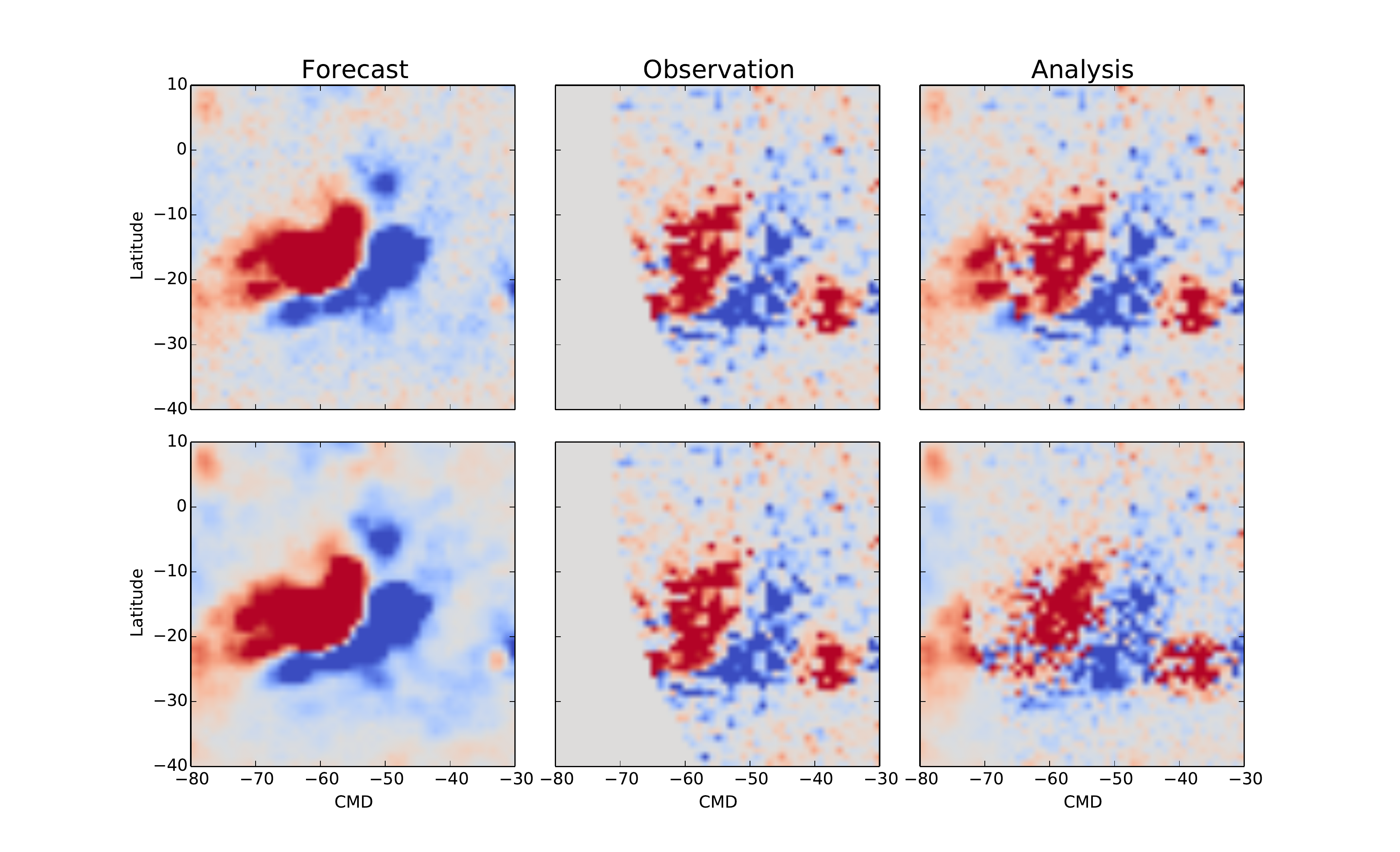}}
  \caption{Comparison of ENLS and LETKF assimilation effects: Top row
    represents one assimilation step using ENLS, while the bottom row utilizes
    LETKF. The first column is the mean forecast, second column is the
    observation, and the last column is the analyzed ADAPT ensemble mean. The
    ENLS algorithm performs a pixel-by-pixel assimilation and therefore
    ignores spatial correlations in the ensemble. This leads to discarding the
    ADAPT ensemble forecast in favor of the observations. Since the LETKF
    takes into account local spatial correlations, large structures, such as
    this active region, are better preserved. The data being assimilated were
    observed from SOLIS-VSM on 26 September 2003 at 16:51.}
  \label{fig:enlsVSletkf_activeregion}
\end{figure}

Overall the result of the ENLS data assimilation is to assign a much greater
weight to the observations than the model. This reduces the information gained
by including the Worden--Harvey model for photospheric flux transport. By
observing one assimilation step, for the same active region portrayed in
Figure \ref{fig:etkfVSletkf_activeregion}, we can see how the ENLS favors the
observed magnetic flux more than the ADAPT ensemble model structure. In Figure
\ref{fig:enlsVSletkf_activeregion} the ENLS data assimilation step drastically
modifies the shape of the active region in its analysis ensemble whereas the
LETKF blends the information from the Worden--Harvey model and the
observations, maintaining the structure of the active region.

\section{Discussion}\label{discussion}

Three different varieties of ensemble Kalman filtering have been compared for
use as data assimilation methodologies in forecasting the global solar
photospheric magnetic flux. Each of these methods treated the flux transport
covariance structure differently, and we showed that this led to drastically
different effects. Although the ensemble Kalman methods have been studied in
the context of solar weather before
\cite{kitiashvili2008application,butala2010dynamic,dikpati2014} the use of a
physical model combined with real solar observations in the application to the
photosphere is new. Moreover, we have highlighted the importance of
considering spatial covariance in such data assimilation schemes and being
wary of relying on a global covariance structure estimated by sampled ensemble
members.

The current ADAPT framework, using the LETKF implementation in the LANL data
assimilation module, does a noticeably better job at balancing the spatial
propagation of information away from the point of observation. Ensemble Kalman
filtering, as opposed to ensemble least squares filtering, also preserves the
variance in the ensemble near data much more. This allows for a more diffuse
model forecast in regions where observations have not yet been made, which
increases the chance of the ensemble range capturing the true photospheric
flux in these regions.

At the same time, a problem in photosphere forecasting noted previously
\cite{yeates2008exploring,mackay2012sun} is loss of balance in magnetic flux
when assimilating large solar active regions on the boundary of the
observation region. When this happens, the observations observe only one
polarity of what should be a coupled polarity active region. Since the WH
model does not include active region creation, for emerging active regions the
ADAPT ensemble cannot have members that include the opposite polarity region
outside of the observation domain \cite{yeates2008exploring}.

In the near future, we plan to incorporate farside estimates of newly emerged
strong magnetic regions (that is, regions not directly observed from the
Earth-side of the Sun) into the ADAPT model. A preliminary example of
utilizing farside data within ADAPT is highlighted by \cite{arge2013modeling}.
Estimates of the magnetic field strength and area of farside active regions
are possible with the helioseismic acoustic holography technique
\cite[e.g.][]{LindseyBraun1997}. The measured helioseismic farside phase
delay values have been parameterized in terms of photospheric magnetic field
strength \cite{Gonz2014}, allowing for an estimation of new solar magnetic
activity on the solar farside while updating the Earth-side of global synoptic
maps. The practical application of the farside data has recently been
discussed with regards to space weather parameter forecasting, for example
solar wind \cite{arge2013modeling}, {\it F}$_{10.7}$
\cite{henney2012forecasting,Gonz2014} and Lyman-$\alpha$ irradiance
\cite{Font2009}.

A further improvement that we plan on pursuing for ADAPT is to incorporate
smooth spatial damping of correlations in the local data assimilation regions
for observations farther from the pixel being analyzed
\cite{gaspari1999construction}. In the current LETKF implementation,
observations on the boundary of the local ellipse have the same weight as
observations over the pixel being analyzed. This is known to cause
discontinuities along the edge of the local data assimilation region
\cite{hunt2007efficient,yang2009weight}. By adding a distance-dependent
weighting to the observations within the local ellipse this problem can be
eliminated.

The ADAPT photospheric forecasting capability continues to improve. This will
lead to more timely boundary conditions for coronal and solar wind models
which drive near Earth space weather forecasting. The data assimilation
portion of the ADAPT framework now has the ability to preserve ensemble
variance near observations as well as far from observations. This enables a
more realistic probable range of predictions. However, to notice this
improvement one must be sure to consider the structure of the entire ensemble
forecast as opposed to only comparing the ensemble mean with observed data.

\section*{Acknowledgments}

  This research was primarily supported by NASA Living With a Star project
  \#NNA13AB92I, ``Data Assimilation for the Integrated Global- Sun
  Model''. Additional support was provided by the Air Force Office of
  Scientific Research project R-3562-14-0, ``Incorporation of Solar Far-Side
  Active Region Data within the Air Force Data Assimilative Photospheric Flux
  Transport (ADAPT) Model''. The photospheric observations used in Figures
  \ref{fig:stddev_image_comparison}, \ref{fig:etkfVSletkf_activeregion}, and
  \ref{fig:enlsVSletkf_activeregion} were provided by SOLIS-VSM. Approved for
  public release: LA-UR-14-27938

\bibliographystyle{plain}
\bibliography{ADAPTletkf}

\end{document}